\documentclass[aps,prl,amsfonts,superscriptaddress,twocolumn,showpacs,nofootinbib]{revtex4-1}  
\usepackage{graphicx}  
\usepackage{amssymb}   
\usepackage{amsmath}
\usepackage{hyperref}
\usepackage{verbatim}
\usepackage{multirow}
\usepackage{float,xcolor}

\begin{document}
\title{Direct measurement of the $^7$Be $L/K$ capture ratio in Ta-based superconducting tunnel junctions}
\author{S.~Fretwell}\affiliation{Department of Physics, Colorado School of Mines, Golden, CO 80401, USA}
\author{K.G.~Leach}\email{kleach@mines.edu}\affiliation{Department of Physics, Colorado School of Mines, Golden, CO 80401, USA}
\author{C.~Bray}\affiliation{Department of Physics, Colorado School of Mines, Golden, CO 80401, USA}
\author{G.B.~Kim}\affiliation{Nuclear and Chemical Sciences Division, Lawrence Livermore National Laboratory, Livermore, CA 94550, USA}
\author{J.~Dilling}\affiliation{TRIUMF, 4004 Wesbrook Mall, Vancouver, BC V6T 2A3, Canada}
\author{A.~Lennarz}\affiliation{TRIUMF, 4004 Wesbrook Mall, Vancouver, BC V6T 2A3, Canada}
\author{X.~Mougeot}\affiliation{CEA, LIST, Laboratoire National Henri Becquerel, CEA-Saclay 91191 Gif-sur-Yvette Cedex, France}
\author{F.~Ponce}\affiliation{Department of Physics, Stanford University, Stanford, CA 94305, USA}\affiliation{Nuclear and Chemical Sciences Division, Lawrence Livermore National Laboratory, Livermore, CA 94550, USA}
\author{C.~Ruiz}\affiliation{TRIUMF, 4004 Wesbrook Mall, Vancouver, BC V6T 2A3, Canada}
\author{J.~Stackhouse}\affiliation{Department of Physics, Colorado School of Mines, Golden, CO 80401, USA}
\author{S.~Friedrich}\affiliation{Nuclear and Chemical Sciences Division, Lawrence Livermore National Laboratory, Livermore, CA 94550, USA}
\date{\today}

\begin{abstract}
We report a high-statistics measurement of the $L/K$ orbital electron capture (EC) ratio in $^7$Be embedded in cryogenic Ta.  The thin Ta film formed part of a high-resolution superconducting tunnel junction (STJ) radiation detector that was used to identify the signals from different decay channels.  The measured $L/K$ capture ratio of 0.070(7) is significantly larger than the only previous measurement of this quantity and the theoretical predictions that include in-medium effects. This value is a uniquely sensitive probe of the 1s and 2s orbital overlaps with the nucleus, and is of relevance to nuclear and atomic physics, as well as Li production in novae and other astrophysical scenarios. This is the first experiment that uses STJs for nuclear-recoil detection, opening a new experimental avenue for low-energy precision measurements with rare isotopes.
\end{abstract}

\maketitle
The vast majority of quantum states in atomic nuclei that have been observed and characterized over the last 80 years result from the detection of radiation emitted in the keV to MeV energy range~\cite{NNDC}.  These measurements have been developed and refined using what are now considered standard experimental techniques~\cite{Knoll}.  For the past 20 years, however, new methods have been developed to perform precision measurements of the low-energy recoiling atom following the emission of these energetic quanta of radiation.  This is typically done in a backing-free environment using trapped radioactive ions~\cite{Sci12,Koz07,Tan14,Del19} or neutral atoms~\cite{Beh97} to perform searches for beyond Standard Model physics~\cite{Yee13,Ste15,Fen18}.  Even with these sensitive techniques, significant systematics still remain due to the nature of environmental control with atom/ion traps.  The recent use of quantum calorimeters can overcome some of these challenges in nuclear physics, albeit at the cost of reduced statistical precision due to the low count rate of these detectors.  Here, we present the first use of high-rate quantum sensors for high-resolution, low-energy nuclear recoil detection in the electron capture (EC) decay of $^7$Be.

The EC decay of $^7$Be is the dominant production mechanism of $^{7}$Li in the core of the Sun, in H-burning layers in Red-Giant Branch (RGB) and Asymptotic Giant Branch (AGB) stars, and in classical nova explosions. Its direct influence on the solar neutrino spectrum has been well known for decades, and heavily studied experimentally in recent years~\cite{Ago18}. Modern microscopic calculations~\cite{Sim13,Ves19} of the electron density at the nucleus have shown that traditional approaches for $^7$Be used in the Solar Standard Model (SSM)~\cite{Bah01} at solar-specific temperature ($T$) and density ($\rho$) cannot reliably be extrapolated to other $T, \rho$ regimes, especially at the lower values seen in RGB and AGB stars. In these highly convective scenarios, the calculated effective $^{7}$Be EC rate using various approaches can differ by an order of magnitude, while the absolute value differs by many orders of magnitude, and thus have important implications for $^{7}$Li survival. In fact, understanding how the $^{7}$Be EC rate changes in dynamic environments is particularly relevant in light of the recent direct detection of $^{7}$Be in the ejecta of several novae through UV and visible observations of $^{7}$Be$^+$~\cite{Mol16,Mol20}, confirming that the majority of Li production in the Universe arises from stellar novae. Here, some observations have inferred an ejected $^{7}$Be mass fraction an order of magnitude larger than theoretical models predict. The dynamical timescale in the expanding ejecta preceding the moment of observation is comparable to the $^{7}$Be half-life, which may itself vary drastically over the $T,\rho$ trajectory due to variations in the EC rate as a function of atomic charge state~\cite{Lit07}. Reliable calculations of the $^{7}$Be lifetime, taking into account both continuum capture and different ionization states over the applicable $T,\rho$ history, are thus needed to rigorously make comparisons between nova models and observations. The theoretical evaluations must therefore account not only for the nuclear interaction, but also the environment in which $^7$Be ions and electrons interact, and require benchmarking to terrestrial conditions where data exist.  In particular, the spatial extent of the $2s$ electronic orbit plays a major role in such estimates~\cite{Das05}, and thus direct measurements of the $L/K$ capture ratio provide important experimental constraints on the atomic properties of the $1s$ and $2s$ wavefunction overlap with the nuclear volume.      

$^7$Be decays by EC primarily to the nuclear ground-state of $^7$Li with a $Q_{EC}$-value of 861.89(7)~keV~\cite{AME16} and a laboratory half-life of $T_{1/2}=53.22(6)$~days.  A small branch of 10.44(4)\% results in the population of a short-lived excited nuclear state in $^7$Li ($T_{1/2}=72.8(20)$~fs)~\cite{Til02} that de-excites via emission of a 477.603(2)~keV $\gamma$-ray~\cite{Hel00}.  In the EC process, the electron can be captured either from the $1s$ shell ($K$-capture) or the $2s$ shell of Be ($L$-capture). For $K$-capture, the binding energy of the $1s$ hole is subsequently liberated by emission of an Auger electron whose energy adds to the decay signal and separates it from the $L$-capture signal.  Since the nuclear decay and subsequent atomic relaxation occur on short time scales (and the emitted neutrino is not detected), a direct measurement produces a spectrum with four peaks: two for $K$-capture and two for $L$-capture into the ground state and the excited state of $^7$Li, respectively.

The current literature value for the $L/K$ capture ratio in $^7$Be of 0.040(6) is based on a single direct measurement at 0.06~K using a high-resolution Si microcalorimeter whose HgTe absorber had been implanted with $^7$Be~\cite{Voy02}.  These measurements were affected by low statistics and an unexpected line-shape of the Doppler-broadened $^7$Li recoil that had to be fit using a range of different assumptions about the recoil slow-down.  At the time of its publication, the measured value was more than a factor of two smaller than the best theoretical estimates.  The authors suggested that this deviation from theoretical predictions resulted from a modification of the electron wavefunction due to in-medium effects, where the Be $2s$ wavefunction was altered by implantation into the HgTe matrix.  This effect was investigated theoretically in detail for a number of host media and was found to decrease the $L/K$ ratio by up to 80\% relative to the value expected in atomic Be~\cite{Ray02}, even if the total decay half-life remained constant to within better than $1\%$.  This reduction has been found to alter the SSM estimates for the $^7$Be and $^8$B neutrino fluxes by $2.0-2.7\%$~\cite{Das05}.  Given the importance of these $L/K$ capture ratios for the extrapolation of $^7$Be EC rates in astrophysical environments, further experimental investigations of these in-medium effects are required.  This Letter reports a high-statistics direct measurement of the $L/K$ capture ratio in $^7$Be implanted in cryogenic Ta where such in-medium effects are predicted to be large.

\begin{figure}[t!]
\centering
\includegraphics[width=0.8\linewidth]{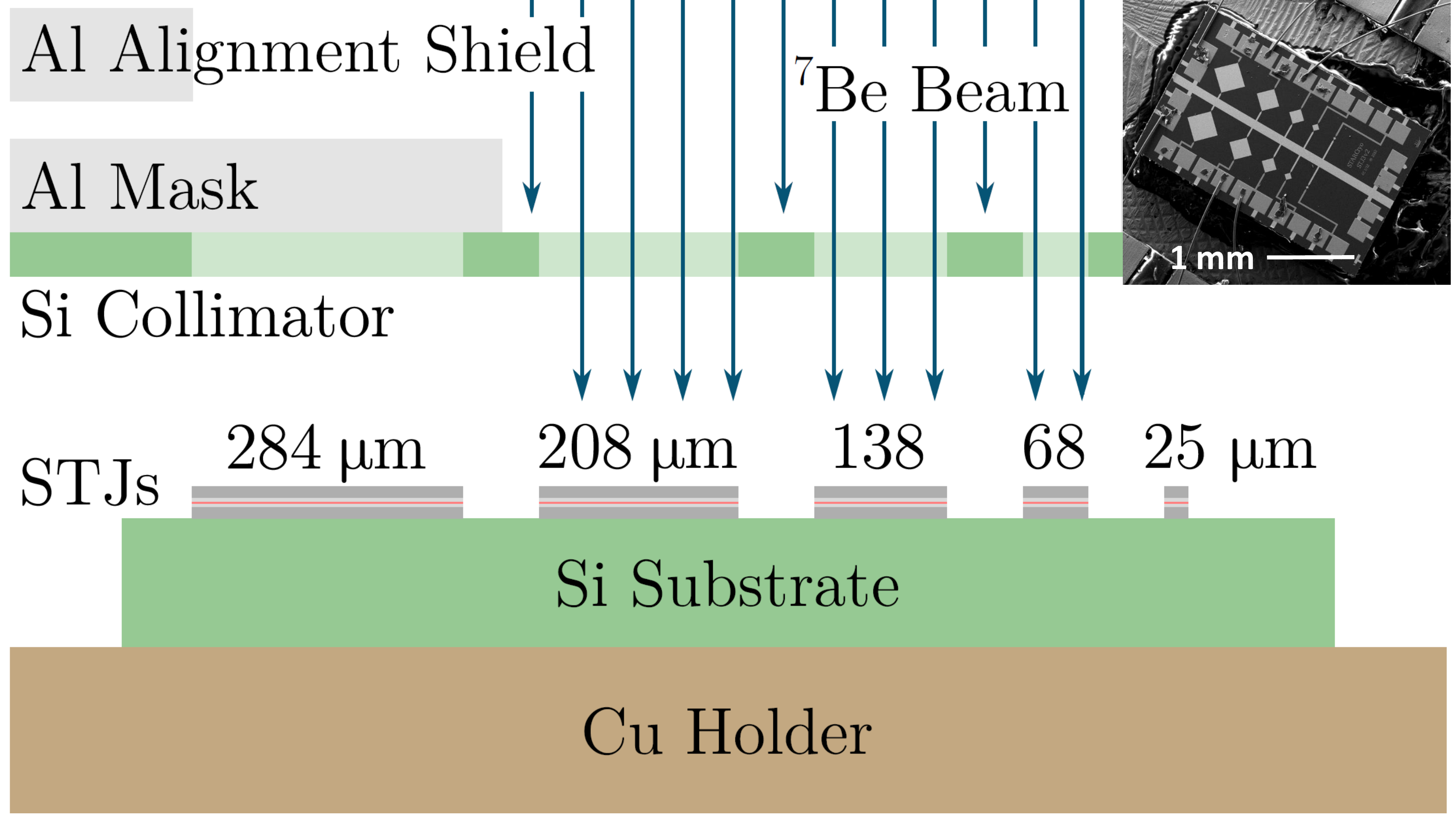}
\caption{\label{BeEST}(Color online) Schematic of the STJ detector chip, Si collimator, and Al mask during $^7$Be implantation. The inset shows a scanning electron microscope (SEM) image of the chip used in this experiment.}
\end{figure}

Ta-based superconducting tunnel junction (STJ) radiation detectors were used to measure the $L/K$ capture ratio of $^7$Be in Ta. The STJs are five-layer devices consisting of Ta (165~nm) - Al (50~nm) - Al$_2$O$_3$ (1~nm) - Al (50~nm) - Ta (265~nm) that were fabricated by photolithography at STAR Cryoelectronics LLC~\cite{Car13} (Fig.~\ref{BeEST} inset).  STJs exploit the small energy gap $\Delta_\mathrm{Ta}=0.7$~meV in superconducting Ta to provide $\sim30\times$ higher energy resolution than conventional Si or Ge detectors, and can resolve the $K$-capture from the $L$-capture signals in the decay.

$^7$Be$^+$ ions were implanted into the STJs though Si collimators at TRIUMF's Isotope Separator and ACcelerator (ISAC) facility in Vancouver, Canada~\cite{Dil14} (Fig.~\ref{BeEST}).  In total, $2\times10^{8}$ radioactive $^7$Be$^+$ ions were implanted into the (138~$\mu$m)$^2$ STJ detector used for the measurements reported here.  The $^7$Be was produced using the isotope separation on-line (ISOL) technique~\cite{Blu13} via spallation reactions from a 10 $\mu$A, 480-MeV proton beam incident on a stack of thin uranium carbide targets. The reaction products were released from the targets, selectively laser ionized~\cite{Las09}, mass-selected, and implanted into the STJs at an energy of 25~keV.  This value was limited by bias-voltage instability of the ISAC target module during the scheduled implantation period and resulted in the ions being implanted closer to the surface than initially desired.  Stopping Range of Ions in Matter (SRIM) simulations~\cite{SRIM} indicate that the mean implantation depth into the Ta layer of each STJ was 46~nm with a straggle of 26~nm.

\begin{figure}[t!]
\includegraphics[width=\linewidth]{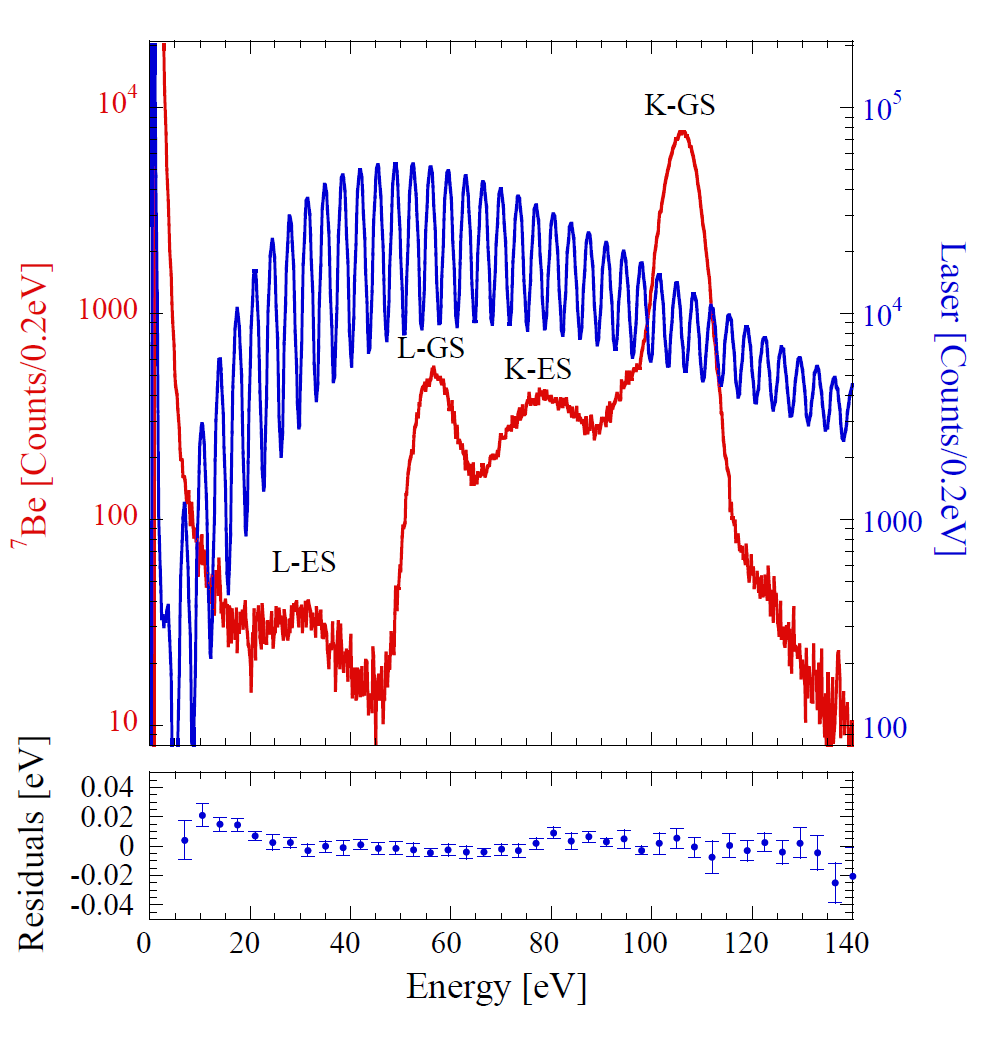}
\caption{\label{CalibrationSpectrum}(Color online) The $^7$Li recoil energy spectrum (red) for a single 22-hour run composed of 44 individually calibrated 30-minute spectra. The calibration signal (blue), whose peaks correspond to integer numbers of laser photons per pulse, was measured in coincidence with the laser trigger and the signal from the $^7$Be decay in anti-coincidence. The residuals in the bottom panel show a calibration accuracy in the region of interest of better than $\pm10$~meV.}
\end{figure}

The decay of $^7$Be was measured at a temperature of $\sim$0.1~K in an adiabatic demagnetization refrigerator (ADR) with liquid N$_2$ and He pre-cooling at Lawrence Livermore National Laboratory (LLNL). For energy calibration, the STJs were simultaneously exposed to 3.49965(15)~eV photons from a pulsed Nd:YVO$_4$ laser triggered at a rate of 100~Hz~\cite{Pon18}.  At this rate, 1\% of the total acquisition time was used for calibration. The laser intensity was adjusted such that multi-photon absorption provides a comb of peaks over the energy range of interest from 20 - 120 eV in the (138~$\mu$m)$^2$ STJ. The signals were read out with a custom-designed current sensitive preamplifier~\cite{War15}, filtered with a spectroscopy amplifier (Ortec 627) with a shaping time of 10~$\mu$s and captured with a two-channel nuclear multi-channel analyzer (MCA) (Ortec Aspec927). The laser calibration spectrum was recorded in coincidence with the laser trigger and the $^7$Be decay signal in anti-coincidence (Fig.~\ref{CalibrationSpectrum}).

Data were acquired for $\sim$20~hours/day over a period of one month. $^7$Be recoil spectra and their corresponding laser calibration were recorded every 30 minutes so that they could be calibrated individually to correct for small drifts in the detector response. For this, the laser signal was fit to a superposition of Gaussian functions, each corresponding to an integer multiple of the single photon energy. The measured Gaussian centroids scale linearly with energy, with a small quadratic non-linearity of order 10$^{-4}$ per eV~\cite{Fri20}. Laser peaks below 20~eV and above 120~eV were omitted from the calibration because they have poor statistics in the individual 30-minute spectra. The calibrated spectra were re-binned to 0.2~eV and summed.  The $^7$Be spectrum and its laser calibration from a single day of data are shown in Fig.~\ref{CalibrationSpectrum}. The laser peaks are well separated, with the energy resolution increasing from 1.4~eV to 2.9~eV full-width at half-maximum (FWHM) in the energy range up to 140~eV. For energies between 20 and 120~eV, the residuals to the calibration have an average uncertainty $\leq$10~meV that sets the calibration accuracy of the spectrum.

To first order, the $^7$Li recoil spectrum is well described by the four peaks generated by the two nuclear and two atomic processes: $K$-capture to the nuclear ground-state (K-GS), $K$-capture to the nuclear excited-state (K-ES), $L$-capture to the nuclear ground-state (L-GS), and $L$-capture to the nuclear excited state (L-ES).  Both decays to the excited nuclear state in $^7$Li display Doppler-broadened recoil peaks due to the isotropic $\gamma$-decay in-flight prior to stopping.  In contrast to previous work, the excited-state peaks are clearly resolved and their line-shape is well described by a Gaussian function, raising the possibility that the Doppler-broadening may have been overestimated in Ref.~\cite{Voy02}.

In addition to the four main peaks, the sudden change in $Z$ following EC decay also results in the electrons not being in an eigenstate of the daughter atom~\cite{Car68}.  This can cause the remaining electron(s) to undergo \textit{shake-up} into a bound state of Li or \textit{shake-off} into the continuum~\cite{Bam77} and generates a high-energy tail above each peak~\cite{Law73}.

The K-GS and L-GS peaks were found to be roughly four times wider than the resolution of the laser peaks.  Detector drift may account for at most $\sim0.5$~eV of this difference.  Other sources of broadening are from variations in the Li $1s$ and $2s$ binding energies for Li atoms at different sites in the Ta lattice, or due to varying self-recombination of the excess quasiparticles in the region of reduced gap due to localized energy deposition of the recoil~\cite{Zeh95}. Further, the energy difference between the L-GS and the K-GS peaks was measured to be 49.27(6)~eV, significantly lower than the binding energy of the Li $1s$ level of 54.74(2)~eV~\cite{Bea67}.  This is either due to in-medium effects of the Li in Ta or an inconsistency in the literature value~\cite{Sie34,Bea67}.  Both the origin of the line broadening and the binding energy discrepancy will be investigated in future work.

In addition, a broad background is visible at low energies that decreases as a function of energy. This background is due to 478~keV $\gamma$-rays from the decay of the $^7$Li excited state that interact in the Si substrate below the STJ detector (Fig.~\ref{BeEST}). High-energy phonons generated in these interactions can propagate to the STJ detectors before thermalization and break Cooper pairs in the detector electrodes. The resulting signals are determined by the deposited energy and the distance between the interaction location and the STJ. A low-energy tail is also visible below the K-GS peak. This tail has different shapes depending on which STJ detector is used, indicating that it is not intrinsic to the $^7$Be decay but due to some form of energy loss through the detector surface which varies for each STJ. Since no tail is observed below the L-GS peak, the low-energy tail below the K-GS peak is likely caused by partial escape of the energy of the Auger electron that is produced in the relaxation of the 1s core hole. This results from the relatively low implantation depth of the $^7$Be atoms causing a significant fraction of the decays to occur within $\sim$10~nm of the detector surface. A similar low-energy tail is expected to also accompany the K-ES peak.  These additional features increase the difficulty in extracting the $L/K$ ratio from the experimental data since the exact shapes of the $\gamma$-ray background and escape tails are not known.

\begin{figure}[t!]
\includegraphics[width=\linewidth]{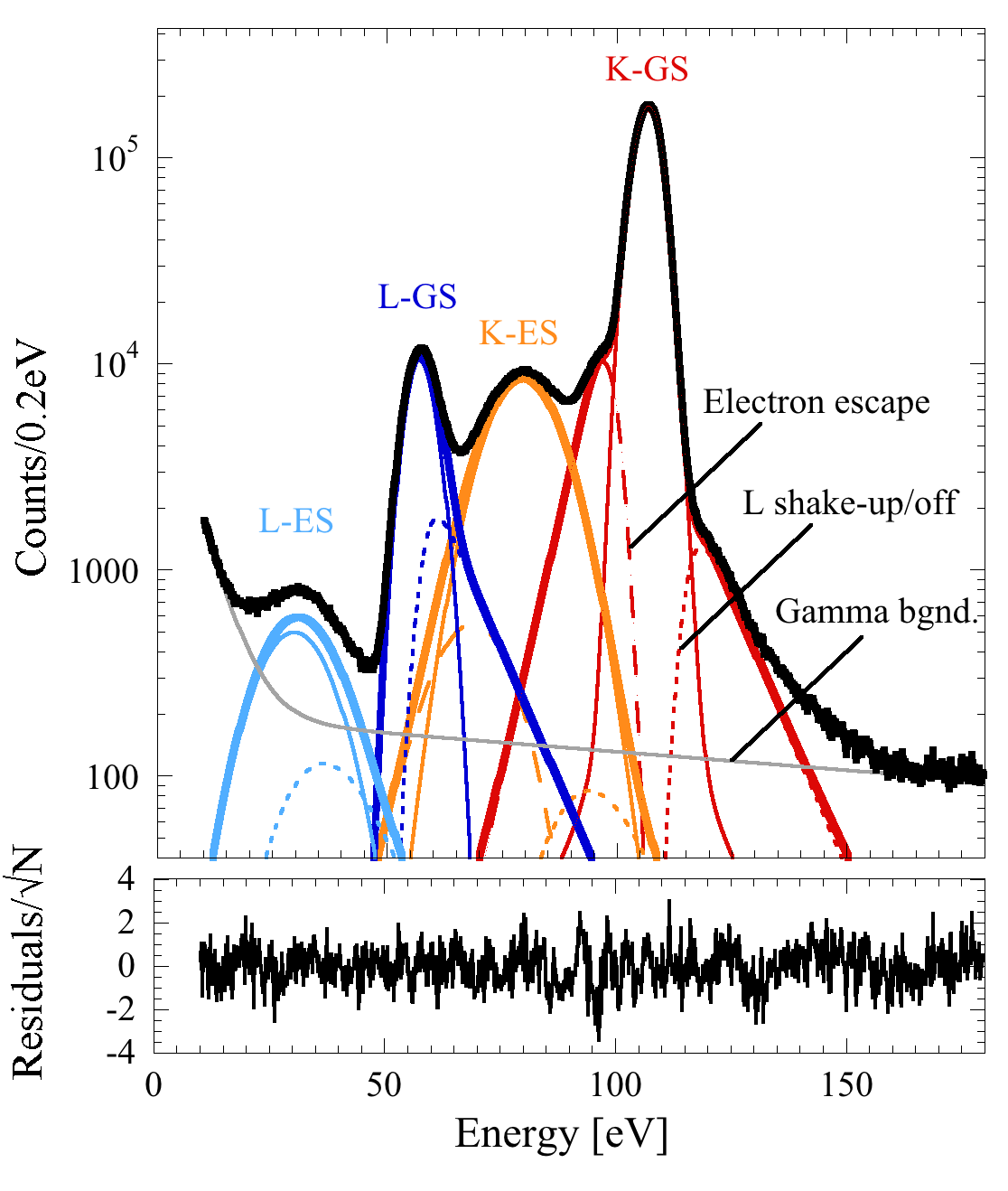}
\caption{\label{BeESTSpectrum}(Color online) The summed $^7$Li recoil spectrum of all individually calibrated 1-day spectra from a single (138~$\mu$m)$^2$ STJ detector. In addition to the four primary peaks (solid lines), the $K$-capture peaks have low-energy tails due to electron escape (dashed lines), and all peaks have a high-energy tail from electron shake-up and shake-off effects (dotted lines). The fit residuals in the bottom panel are normalized by the statistical error of the measurement ($\chi^2/\nu=0.95$).}
\end{figure}

The resulting recoil-energy spectrum from all runs, energy calibrated and summed, is shown in Fig.~\ref{BeESTSpectrum}.  For the two ground-state decay peaks, three Gaussian functions were used to fit the edges of the distributions, which may reflect the broadening of the peaks due to in-medium effects.  The two excited-state peaks are well described by the same set of Gaussian distributions but with a Doppler-broadened width.  All $K$-capture Gaussian functions were convolved with the Lorentzian lineshape of the Auger process defined by the lifetime of the Li $1s$ core hole~\cite{Cit77}.  To aid in the fit convergence, the known branching fraction of EC decay to the nuclear excited-state in $^7$Li~\cite{Til02} was used to constrain the area ratios of the excited-state to ground-state for both $K$- and $L$-capture.  The $L$-electron shake-up and shake-off tails for all four peaks were approximated by an exponential decay convolved with the width of the corresponding peak~\cite{Law73}. The tail areas and decay constants were constrained to the same value respectively for the two $K$- and the two $L$-capture peaks.  Similarly, the electron escape following Auger emission was modeled by exponentially-decaying tails below the $K$-capture peaks, again broadened by the widths of the K-GS and K-ES peak, respectively.  Finally, the $\gamma$-ray background was described by a sum of two exponential functions based on spectra from the STJ detectors without $^7$Be, as well as previous work with the same detectors~\cite{Pon18}.

The fit was performed on the data from 10 to 180~eV with a least-squares regression within the Python \textsc{iminuit} framework~\cite{iminuit,Jam75}.  The best fit is presented with the data in Fig.~\ref{BeESTSpectrum}, and resulted in $L/K=0.07165(29)_\textrm{stat.}$ with $\chi^2/\nu=0.95$.  To investigate the systematic uncertainties in the $L/K$ ratio that result from the choice in peak shape and exponential tails, the functions and constraints of the fit were systematically varied over a range of plausible alternatives.  This included using two instead of three Gaussian functions for the central peaks, both with and without steep exponential tails.  More importantly, different constraints on the areas of the shake-up/shake-off tails were included, because when left unconstrained, all fits converged to a larger tail above the $L$-capture peaks than above the $K$-capture peaks (Fig.~\ref{BeESTSpectrum}).

For $K$-capture, $L$-electron shaking was found to be between 1.1\% and 2.8\%, while for $L$-capture their contribution varied from 1.2\% to 31.5\%.  It is this difference that ultimately dominates the systematic uncertainty in the $L/K$ ratio.  These variations could simply reflect that the variation in the Li $2s$ energy levels in the Ta matrix is larger than the spread of the Li $1s$ levels, although a distribution of well above 10~eV does seem unrealistically wide.  Alternatively, the $L$ shake-up and shake-off effects could be stronger for $L$-capture than for $K$-capture events, although this is in disagreement with theoretical estimates and thus also unlikely. Finally, the 60-70~eV region could be affected by an electron escape tail that does not decay exponentially to zero but has an increased probability for high energies. This could reflect that a fraction of Auger electrons escape from the detector with almost all of their energy since $\leq1$\% of the $^7$Be atoms were implanted within the mean free path of $\sim1$~nm of the surface~\cite{Zia06}.  The different fits resulted in a spread of the $L/K$ ratio of 0.064 to 0.078 with reduced $\chi^2$ values between 0.95 and 1.41.  This range is dominated by the uncertainty of the fit in the region from 60 to 70 eV and reflects $L/K$ ratios that are consistent with the data for a variety of assumptions in the underlying physics.  Taking into account these differences, we determine a value of $L/K=0.0702(66)_\mathrm{sys.}(3)_\mathrm{stat.}$.

The measured value of $L/K=0.070(7)$ is nearly a factor of 2 larger than the only previous measurement of $0.040(4)$ in HgTe~\cite{Voy02}.  This discrepancy likely results from the lower statistics of the earlier measurement and the associated uncertainty of the fit functions, as well as the different host materials.  Since the quantitative origin for this large discrepancy cannot be determined, we recommend the value of $L/K=0.070(7)$ for future modeling of the $^7$Be/$^7$Li system.  We also performed a precise new calculation of the $L/K$ ratio for atomic $^7$Be decay in-vacuum using the method detailed in Refs.~\cite{Mou18,Mou19}, and yields a value of $L/K=0.105(8)$.  For $^7$Be in Ta, in-medium effects are expected to change this value by a factor of 0.2986~\cite{Ray02} to $L/K = 0.031(2)$.  This is more than 2 times ($4\sigma$) smaller than the measured value presented here.  The discrepancy between the theoretical estimates for the capture ratio in $^7$Be and the experimental result likely points to a deficiency in the models that are used to correct for in-medium effects.  Since the measured $L/K$ ratio in $^7$Be is sensitive to the $1s$ and $2s$ wavefunction overlap with the nuclear volume, they are an important empirical input for modeling in several research areas, particularly EC rates in any astrophysical scenario that deviates significantly from conditions the core of the sun, where such effects are not yet included.  Finally, these measurements demonstrate that high-resolution STJ detectors can be used for low-energy nuclear recoil detection, and show tremendous promise for high-rate quantum sensing experiments in subatomic physics.

\begin{acknowledgments}
This work was funded by the LLNL LDRD program through grants 19-FS-027 and 20-LW-006 and the U.S. Department of Energy, Office of Science under contract DE-SC0017649.  TRIUMF receives federal funding via a contribution agreement with the National Research Council of Canada (NRC).  This work was performed under the auspices of the U.S. Department of Energy by Lawrence Livermore National Laboratory under Contract DE-AC52-07NA27344.  The theoretical work was performed as part of the EMPIR project 17FUN02 MetroMMC. This project has received funding from the EMPIR programme co-financed by the participating states and from the European Union's Horizon 2020 research and innovation programme.  We would like to thank Friedhelm Ames, Louis Clark, Peter Kunz, Jens Lassen, and Brad Schultz for their efforts in facilitating the ion-beam implantation.  KGL also thanks Tibor Kibedi, Uwe Greife, John Behr, Fred Sarazin, Vladan Stefanovic, Mark Lusk, Xerxes Stirer, and Jeramy Zimmerman for useful discussions.
\end{acknowledgments}

\bibliography{references}

\end{document}